 \documentclass[sigconf]{acmart}
\usepackage{array}
\AtBeginDocument{%
  }

\copyrightyear{2025}
\acmYear{2025}
\setcopyright{rightsretained}
\acmConference[EduCHI '25]{EduCHI 2025: 7th Annual Symposium on HCI Education}{July 30-August 01, 2025}{Bloomington, IN, USA}
\acmBooktitle{EduCHI 2025: 7th Annual Symposium on HCI Education (EduCHI '25), July 30-August 01, 2025, Bloomington, IN, USA}
\acmPrice{}
\acmDOI{10.1145/3742901.3742909}
\acmISBN{979-8-4007-1463-4/25/07}

\begin{document}




\title{Rethinking Citation of AI Sources in Student-AI Collaboration within HCI Design Education}


\author{Prakash Shukla}
\email{shukla37@purdue.edu}
\orcid{0009-0002-7416-1758}
\affiliation{
  \institution{Purdue University}
  \city{West Lafayette}
  \state{Indiana}
  \country{USA}
}
\author{Suchismita Naik}
\email{naik33@purdue.edu}
\orcid{0009-0002-5667-4576}
\affiliation{
  \institution{Purdue University}
  \city{West Lafayette}
  \state{Indiana}
  \country{USA}
}
\author{Ike Obi}
\email{obii@purdue.edu}
\orcid{0000-0002-3910-8890}
\affiliation{
  \institution{Purdue University}
  \city{West Lafayette}
  \state{Indiana}
  \country{USA}
}
\author{Jessica Backus}
\email{backus@purdue.edu}
\orcid{0009-0008-0863-8080}
\affiliation{
  \institution{Purdue University}
  \city{West Lafayette}
  \state{Indiana}
  \country{USA}
}
\author{Nancy Rasche}
\email{nrasche@purdue.edu}
\orcid{0000-0002-8042-9258}
\affiliation{
  \institution{Purdue University}
  \city{West Lafayette}
  \state{Indiana}
  \country{USA}
}
\author{Paul Parsons}
\email{parsonsp@purdue.edu}
\orcid{0000-0002-4179-9686}
\affiliation{
  \institution{Purdue University}
  \city{West Lafayette}
  \state{Indiana}
  \country{USA}
}

\renewcommand{\shortauthors}{Shukla et al.}

\begin{abstract}
  The growing integration of AI tools in student design projects presents an unresolved challenge in HCI education: how should AI-generated content be cited and documented? Traditional citation frameworks—grounded in credibility, retrievability, and authorship—struggle to accommodate the dynamic and ephemeral nature of AI outputs. In this paper, we examine how undergraduate students in a UX design course approached AI usage and citation when given the freedom to integrate generative tools into their design process. Through qualitative analysis of 35 team projects and reflections from 175 students, we identify varied citation practices ranging from formal attribution to indirect or absent acknowledgment. These inconsistencies reveal gaps in existing frameworks and raise  questions about authorship, assessment, and pedagogical transparency. We argue for rethinking AI citation as a reflective and pedagogical practice; one that supports metacognitive engagement by prompting students to critically evaluate how and why they used AI throughout the design process. We propose alternative strategies—such as AI contribution statements and process-aware citation models that better align with the iterative and reflective nature of design education. This work invites educators to reconsider how citation practices can support meaningful student–AI collaboration.
\end{abstract}




\maketitle

\section{Introduction}

Artificial intelligence (AI) tools have increasingly become embedded within the design processes of students, transforming traditional methods of ideation, prototyping, and evaluation \cite{lively_integrating_2023, chellappa_understanding_2024, papachristos_integrating_2024}. As these tools become commonplace in educational settings, understanding how students articulate their use---and the rationale behind incorporating AI---is crucial for both pedagogical and ethical reasons. Proper citation and clear descriptions of AI tool integration in documentation not only support transparency but also enhance the reflective practices of design students regarding their design decisions \cite{chen_probing_2021}. Central to design education is the practice of clearly articulating and justifying design decisions through structured argumentation. Design argumentation involves communicating the rationale behind design choices, explicitly connecting theoretical, empirical, and material grounds to support design claims \cite{dalsgaard_design_2013}. Providing design argumentation helps students to critically examine and communicate their design processes \cite{dalsgaard_design_2013}, and effective design argumentation enables students to convincingly ``sell'' their work, making their decision-making process transparent to stakeholders such as instructors and peers \cite{gray_narrative_2018}.

The ability to clearly articulate design rationale holds significant pedagogical value. Instructors rely on design argumentation to assess not only final solutions, but also students' underlying thought processes and methods. Engaging students in design argumentation cultivates a shared vocabulary and sensitivity to design values \cite{dalsgaard_design_2013}. Feedback and critique, central to studio-based pedagogy, explicitly foreground design rationale and create discursive spaces where students' decisions can be challenged and refined, providing important learning opportunities \cite{dannels_performing_2005}. Clear articulation of AI tool usage within design arguments thus allows instructors valuable insights into the design thinking processes of the students, the quality of their engagement with design methods, and the overall rigor of their approach. 

As AI becomes part of the design toolkit of students, design argumentation must now include AI-generated contributions alongside the design judgment of the students. In this context, argumentation and citation become deeply intertwined: citing AI use is not just about attribution—it helps make the reasoning behind design decisions visible. When students engage AI across stages like problem framing, ideation, wireframing, and evaluation, citing those interactions becomes essential to surface the collaborative nature of the process. Without such documentation, key aspects of student reasoning may be obscured, complicating assessment and limiting reflective learning. This tension calls for a closer look at whether existing citation frameworks adequately capture the evolving dynamics of student–AI collaboration and support educators in tracing design cognition in AI-mediated workflows.

Traditional citation frameworks were developed for stable, authored, and retrievable sources, allowing educators to trace the intellectual influences of the design process of the students and assess their design cognition \cite{lanning_modern_2016}. However, as students increasingly integrate generative AI tools into their workflows, these frameworks fall short, introducing risks to both pedagogical clarity and assessment validity. The dynamic, non-retrievable, and context-dependent nature of AI-generated content—such as outputs from ChatGPT, Claude, or Midjourney—raises critical questions: How should students document their collaboration with AI? What counts as meaningful attribution? And how can educators distinguish between student-generated insight and machine-assisted output? Through a case study of 35 HCI student teams (175 students) in a UX design course, we examine how students currently cite—or fail to cite—AI in their design projects. This paper offers a series of provocations around the limitations of existing citation norms and the need to reimagine AI attribution as a pedagogical tool that supports transparency, reflection, and integrity in design education. These provocations include:

\begin{itemize} 
    \item How might we re-imagine citation practices to better reflect the evolving reality of student–AI collaboration in design education? 
    \item Should AI citation be treated as a reflective, pedagogical instrument or merely an academic obligation? 
    \item How can educators balance the need for transparency with the potential stigma students may face when disclosing AI usage? 
    \item What new models, such as AI contribution statements or phase-based documentation, might more accurately represent the role of AI in the design process? 
\end{itemize}

\section{Background}
\subsection{Tension and Dilemma of Citing AI Sources}
Traditional citation practices are built on three foundational principles: credibility, retrievability, and accountability \cite{lanning_modern_2016}. These practices were designed to ensure that human-generated content can be easily attributed to the author, utilizing stable sources that allow for tracing the provenance of the given article or content as it evolves or improves over time. However, the current era of AI-generated content challenges these established norms, as content generated by AI systems is often dynamic and contextual, typically lacking stable reference points \cite{alea_albada_giving_2024}. Even when the reference points are stable, there is no guarantee that all content will be exactly replicated or retrieved, even when using identical prompts \cite{cotton_chatting_2024}. This tension about the suitability of old citation practices in the current era of AI raises important questions about how students should acknowledge the role of AI in their design-focused projects where process documentation is pedagogically significant.

In response to the growing use of AI in student formation, style guides like APA and MLA have begun to offer provisional recommendations for citing AI-generated content \cite{mcadoo_how_2023, istrate_how_2023}. These guidelines typically suggest treating AI as a type of software, focusing on citing the model name, version, and date of access. However, this approach, while pragmatic, falls short of fully capturing the nuanced role AI plays within the design process and timeline. Similarly, universities are developing institutional policies and library guides to address the use of AI in academic work. However, their efforts, while well-intentioned, often approach AI citation through existing frameworks rather than reconceptualizing the citation of AI as a tool of collaboration and co-authoring \cite{chan_generative_2024}. Sandhaus et al. \cite{sandhaus_student_2024} highlighted that institutional responses frequently emphasize compliance over pedagogical value, leading them to treat AI citation primarily as a matter of academic integrity concern rather than an opportunity to document evolving design cognition.


Overall, these challenges present a complex dilemma for students integrating AI into their academic work, particularly in design education, where process documentation and the clear articulation of design decisions are crucial pedagogical components. This dilemma could lead to hesitation in citing and disclosing AI use \cite{wang_they_2024} due to concerns about how educators might perceive excessive usage of AI as a collaborator, potentially leading them to either omit or minimize the impact of AI on their design project in their documentation. Hence, this lack of a forward-looking outlook on AI citation means that, at present, there remains an unresolved tension necessitated by the absence of clear citation guidelines that delineate AI as a collaborator deserving of co-authorship and AI as a tool requiring citation \cite{hosseini_ethics_2023} that impacts the pedagogical experience of HCI students.

\subsection{AI Citation in Design and HCI Education}

Design and HCI education present unique challenges for the current AI citation practices proposed by regulatory bodies due to its emphasis on process, iteration, and design rationale. Within HCI/Design, citation serves not only to attribute sources but also to document the evolution of design thinking \cite{parsons_developing_2023}. Hence, as AI becomes deeply embedded in design processes, from ideation to evaluation, the limitations of proposed AI citation approaches become increasingly glaring. To this end, HCI educators are increasingly highlighting the need for new citing frameworks that capture the dynamic relationship between designers and AI \cite{shi_understanding_2023,lee_when_2025}. Shi et al. \cite{shi_understanding_2023} identified various collaboration patterns between designers and AI, noting that documentation practices have not kept pace with rapidly evolving collaborative workflows. Similarly, Lee et al. \cite{lee_when_2025} examined when and how designers integrate AI into their processes, finding that current documentation practices often fail to capture the nuanced ways AI shapes design decisions.

In the context of HCI education and pedagogy, Parsons and Gray \cite{parsons_separating_2022} emphasize how feedback and assessment in design studios rely on transparency about process and influences. This means that when AI tools become part of the design process, citation practices must evolve to ensure instructors can adequately evaluate student work and provide appropriate guidance. This is because without clear frameworks for documenting AI contributions within design projects, both assessment validity and pedagogical effectiveness may be compromised. Furthermore, the challenge of citing AI in design education thus extends beyond attribution to encompass fundamental questions about design cognition, authorship, and assessment. In addition, Shukla et al. \cite{shukla_-skilling_2025} argue that there is a growing concern about the potential for AI to obscure rather than clarify the design thinking processes of students, particularly when citation practices fail to adequately document the boundaries between student and AI contributions. As a result of these, design educators face a unique challenge that involves developing citation frameworks that not only attribute AI contributions but also preserve the visibility of the design thinking and decision-making process of the student to avoid obscuring the essential aspects of the design cognition of the students and to also make it easy for HCI educators to provide meaningful guidance and evaluation. These have implications beyond what they are doing in the school and would influence the metacognitive skills they would transfer to industry design practice and would be helpful for designers to communicate their ideas effectively to their audience when they transition to industry practitioners \cite{shukla_communication_2024}.

\section{Method: A Case Study of How HCI Students Are Currently Struggling with Citing AI Sources}

In this provocation case study, we explored the AI citation practices of HCI/Design students during design projects, and through this lens, foregrounded tensions, challenges, and opportunities for support in the context of using and reporting AI usage within design projects.

\subsection{Research Context}

We conducted this case study within an undergraduate course in UX Design fundamentals at a large public Midwestern university in the United States. Conducting this case study enabled us to investigate in-depth the AI citation practices of design students within a course setting where design documentation and rationale are central to the learning process and outcome. We designed the study to observe naturalistic student behaviors around AI tool usage and, therefore, provided minimal prescriptions regarding citation formats while explicitly encouraging transparency about the use of AI. This approach enabled us to observe the emergent citation practices of the students in the absence of guidelines rather than enforcing a predetermined citation structure, and thus provided us with insights into the intuitive approaches students adopt to resolve AI attribution in their design projects. Our home university provided guidelines for AI policy in our course subject to the discretion of the instructors. In this project we used our discretion to suspend the application of the policy to this project. This study was approved by the IRB office at our home institution.

\subsection{Participants and Setting}

The study involved a total of 175 students organized into 35 project groups across four sections of the UX Design fundamentals course. The course is a required component in the School's curriculum, typically taken by first and second-year undergraduate students majoring in Design, Computing, Games, Analytics, and related fields. The students had different levels of experience and exposure to UX and design projects in general. Students received basic training in literature review and APA citation practices before the project.

The four course sections were taught by three instructors who coordinated closely through weekly meetings and shared teaching materials to ensure uniformity in instructional content, project guidelines, and assessment criteria across all sections. This coordination was crucial in minimizing instructor-specific influences on how the students approach AI tool usage and citation practices. All instructors followed the same project briefing protocol and  instructional requirements regarding AI usage permissions and documentation expectations. 

\subsection{Data Collection}

The students received a design project prompt that involved designing a \textit{fitness tracking experience}, which was followed by a detailed prompt that explicitly addressed AI usage as follows:

\begin{quote}
    \textit{USE OF GENERATIVE AI. In this project, you are encouraged to leverage Generative AI tools, including LLMs, in your design process. \textbf{Be sure to justify and provide a rationale for their use; give credit to the AI tools with a citation and explanation of how you used the tool}. Include a group reflection (300-500 words) in your documentation discussing your experiences of using these tools in your project, highlighting both their advantages and challenges. Your use of Generative AI and the contents of your group reflection will not negatively impact your Project 3 scores, and for this project, you’ll be exempt from the AI usage policy.}
\end{quote}

The AI usage prompt was carefully designed to create a permissive environment for integrating AI into design projects while emphasizing the need for documentation and reflection. Importantly, the students were notified that the use of AI tools was optional rather than mandatory, allowing us to observe varied approaches to AI integration and reporting. 

The materials collected for analysis included complete project documentation from each team, containing research findings, design process, group reflections (300-500 words), and final design artifacts. Supplementary materials included appendices with AI prompts or conversation logs. In preparation for data analysis, we anonymized the data using a systematic labeling approach: each team was assigned a unique identifier based on the order and frequency of AI tool citations. For instance, ``T20a'' denoted the first reference to AI-generated content made by Team 20 in their documentation, while ``T20b'' indicated their second reference. This nomenclature enabled consistent tracking of AI citations, regardless of content type (e.g., text, image).



\subsection{Data analysis}

\subsubsection{Exploratory Analysis}
We employed qualitative content analysis \cite{elo_qualitative_2008,hsieh_three_2005,forman_qualitative_2007} paired with citation guidelines \cite{borg_citation_2000,fazilatfar_investigation_2018} as our analysis framework to examine the design documentation and presentation slides of the students for their AI citation practices. Before conducting the main analysis, we conducted an inductive thematic analysis to identify emerging patterns from the reflections of the students on the use of AI within their design projects. This exploratory process began with open coding, where each researcher independently reviewed a subset of the student reflections to identify emerging themes related to the use of AI and the rationale the students provided to support their positions. Through iterative discussion and refinement of the themes that emerged, we developed a codebook that supported our main qualitative content analysis.

\subsubsection{Codebook Components}
Our codebook consisted of three dimensions. First, content type, which captured whether AI was used for text (e.g., interview questions, documentation, analysis, scripts) or image generation (e.g., icons, mockups, interface elements). Second, design phase, which tracked when AI was used in the process; problem framing, ideation, prototyping, documentation, or support activities such as interview protocols and video scripting. Third, citation classification, which categorized how students reported AI use: 1) explicit citations with tool name, version, link, in-text and reference entry; 2) general mention without version or date; 3) indirect mention in reflections or appendices only; and 4) no citation at all. This framework supported consistent coding across all submitted projects.

\subsubsection{Main Analysis}

We employed the codebook to conduct a more structured content analysis, \textbf{\textit{focusing specifically on citation practices for this analysis}}. The documentation for each team was systematically examined to extract all instances of reference to AI within the documentation. These instances were then coded using our analytical framework by mapping both the design phase and the citation approach.

\begin{figure}
    \centering
    \includegraphics[width=1\linewidth]{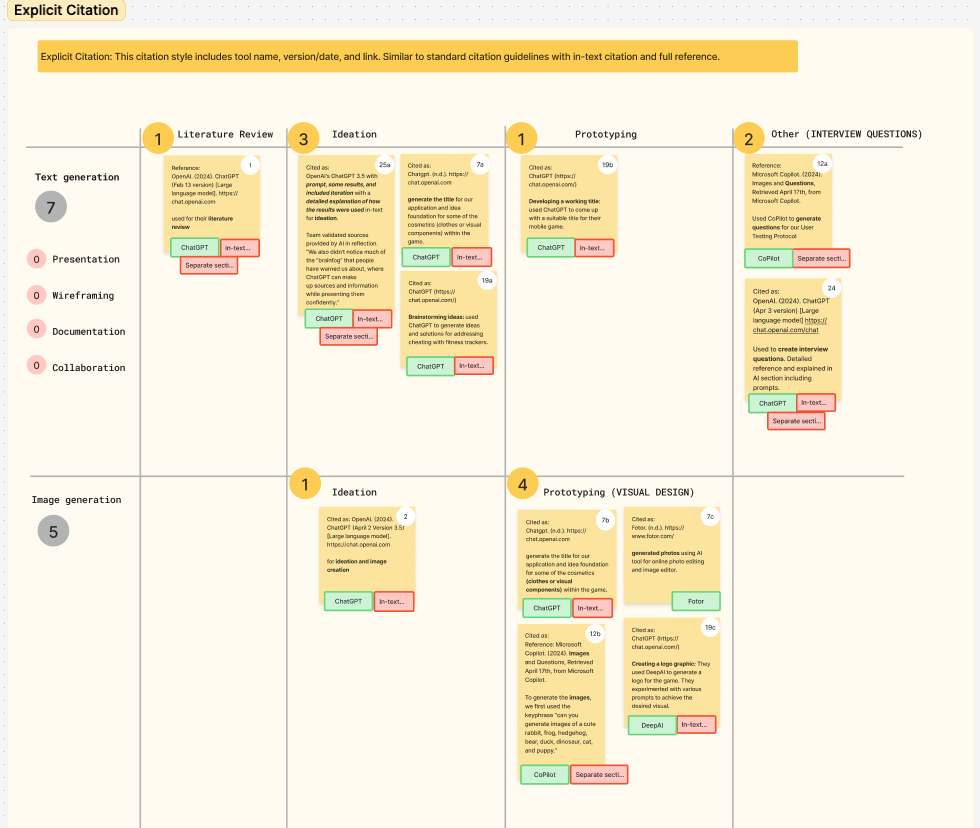}
    \caption{Example of explicit citation showcasing how the framework was used for analysing pattern of AI usage and citation}
    \label{fig:framework_citation}
\end{figure}

Five researchers, including the three course instructors and two faculty members, conducted this analysis to identify patterns in the student reflections and documentation. 
The group reflection section proved especially rich in detailing which AI tools were used, how they were integrated into various design phases, and the perceived advantages and challenges of their usage. We systematically extracted citation practices by closely reading the full documentation of each team. We then organized this data using our framework that mapped AI usage along two dimensions: content type (e.g., text generation, image generation) and design process phase (e.g., literature review, ideation, presentation, wireframing, prototyping, documentation, collaboration, and other activities). This framework is illustrated in Figure~\ref{fig:framework_citation} for the explicit citation theme. We employed Figjam as our collaborative content analysis tool to allow the team to collaboratively iterate and refine our analysis.  To ensure consistency and reliability, the research team held regular meetings and collaborative analysis sessions to validate and verify the emerging patterns and themes, as well as the application of our codebook, including redefining coding definitions where necessary. 

\subsection{Positionality statement:} Five researchers contributed to this study, each bringing distinct disciplinary and cultural perspectives paired with their qualitative research experience. The team represented fields including Design Education, Human-Computer Interaction (HCI), and Computer Science. The two faculty members have a combined more than thirty years of experience teaching design at both undergraduate and graduate levels at higher education institutions across the US and Canada. 
These professional backgrounds and personal experiences informed the research design, analysis process, and interpretation of findings.



\begin{table}[]
    \caption{Distribution of Citation Practices Across Different Functional Applications}
    \centering
    \begin{tabular}{p{0.25\linewidth} p{0.2\linewidth} p{0.25\linewidth} p{0.2\linewidth}} 
        \hline
        \textbf{Citation Category} & \textbf{For Text Generation} & \textbf{For Image Generation} & \textbf{Total (Count,\%)} \\
        \hline
        Explicit Citation & 07 & 05 & 12 (17.14\%) \\
        
        General Mention & 31 & 05 & 36 (51.43\%) \\
        
        Indirect Mention & 16 & 06 & 22 (31.43\%) \\
        \hline
        \textbf{Total} & \textbf{54 (77.14\%)} & \textbf{16 (22.86\%)} & \textbf{70 (100\%)} \\
        \hline
    \end{tabular}
    \label{tab:AI_citation_type}
\end{table}

\section{Findings}
The initial coding process yielded 75 distinct codes derived from the student documentation data. These codes were organized into four overarching themes: Explicit Citation (12), General Mention (36), Indirect Mention or Lack of Explicit Citation (22), and No Citation Mentioned (5). For subsequent content analysis focused on citation practices in relation to content type and design phases, we excluded the five codes falling under ``No Citation Mentioned'', as they lacked sufficient reference detail. The remaining 70 codes were further categorized according to the design phase in which the AI tools were used. These included Literature Review (11), Ideation (21), Prototyping (12), Documentation (11), and Other Activities (15), with the ideation phase representing the highest phase of usage within the design process (30\%). Table ~\ref{tab:AI_usage_design_process} shows this information in detail. Notably, there was no reported use of AI for collaboration, wireframing, or presentation activities. In the following sections, we elaborate on these themes and foreground the various ways the students approached AI citation.

\subsubsection{AI usage across design process} 


\begin{table}[]
    \caption{Frequency of Citations for AI Tool Use Across Design Process Phases and Functional Applications}
    \centering
    \begin{tabular}{p{0.25\linewidth} p{0.2\linewidth} p{0.25\linewidth} p{0.2\linewidth}} 
        \hline
        \textbf{Design Phase} & \textbf{For Text Generation} & \textbf{For Image Generation} & \textbf{Total (Count,\%)} \\
        \hline
        Literature Review & 11 & 0 & 11 (15.71\%) \\
        
        Ideation & 15 & 06 & 21 (30.00\%) \\
        
        Prototyping & 02 & 10 & 12 (17.14\%) \\
        
        Documentation & 11 & 0 & 11 (15.71\%) \\
        
        Others - Interview Questions & 13 & 0 & 13 (18.57\%) \\
       
        Others - Video script & 02 & 0 & 02 (2.86\%) \\
        \hline
        \textbf{Total} & \textbf{54 (77.14\%)} & \textbf{16 (22.86\%)} & \textbf{70 (100\%)} \\
        \hline
    \end{tabular}
    \label{tab:AI_usage_design_process}
\end{table}

\begin{table*}[t]
  \caption{AI Tool usage across design phases during the design project process of the students}
  \label{tab:ai-usecases-design}
  \begin{tabular}{
    p{0.20\linewidth} 
    p{0.35\linewidth} 
    p{0.35\linewidth}
  }
    \toprule
    \textbf{Design Phase} & \textbf{AI Use Case} & \textbf{Tool(s) Used} \\
    \midrule

    \textbf{Ideation} & 
    Title generation, brainstorming, critique, content tone design & 
    ChatGPT, Adobe Firefly \\

    \midrule
    \textbf{Literature Review} & 
    Secondary research, framing problem space, information synthesis & 
    ChatGPT, Claude 3 \\

    \midrule
    \textbf{Prototyping (Text)} & 
    Naming, branding, foundational structure development & 
    ChatGPT \\

    \midrule
    \textbf{Prototyping (Image)} & 
    Logo, icon, mockup, and character image generation & 
    DeepAI, Fotor, Brand Crowd, Uizard AI, Microsoft Copilot, Chroma, ChatGPT (image), Adobe Firefly \\

    \midrule
    \textbf{Documentation} & 
    Drafting, summarizing, rewording, organizing, citation formatting & 
    ChatGPT \\

    \midrule
    \textbf{Others - Interview Questions} & 
    Interview and survey protocol generation, refinement, insight extraction & 
    ChatGPT, Microsoft Copilot, Gemini \\

    \midrule
    \textbf{Others - Video Scripting} & 
    Video scriptwriting for final presentations & 
    ChatGPT \\

    \bottomrule
  \end{tabular}
\end{table*}

AI tools were most prominently used during the ideation phase, where students employed LLMs for brainstorming, concept development, and critique (see Table~\ref{tab:ai-usecases-design}). T19a noted using ChatGPT to \textit{``generate ideas and solutions for addressing cheating with fitness trackers,''} while T13b shared they \textit{``asked AI to critique ideas,''} indicating exploratory and evaluative use. T9c used ChatGPT for user engagement, \textit{``generating sassy and humorous prompts''} and defining app features like \textit{``rewards and penalties and creating engaging scenarios.''} In visual ideation, tools like Adobe Firefly supported early concept sketching and logo generation. T31d reflected that AI helped them \textit{``explore creative concepts and solutions,''} highlighting AI's role in shaping both content and design direction.

During the literature review phase, students used AI tools primarily for information synthesis and problem framing. T9a noted that \textit{``ChatGPT helped define the problem space and conduct secondary research,''} while T13a used it for \textit{``shaping initial framing of the project... scope and target audience.''} T16a provided a prompt—\textit{``Brainstorm and list as many reasons as to why people don't want to or can't use fitness trackers''}—showing how prompt construction became part of the research process, reflecting early metacognitive engagement with AI.


AI-supported prototyping was facilitated through both text and image generation tools. Text-based tools, particularly ChatGPT, were used by several teams to aid conceptual development. For instance, T19b used ChatGPT to generate a title for their mobile game. T26b reported using ChatGPT to establish the foundational elements of their design process, indicating a broader application in defining project structure and direction. 
AI-generated imagery played a prominent role in visual prototyping, with teams leveraging various platforms to produce icons, mockups, logos, and aesthetic components. For instance, T12b used Microsoft Co-pilot for character illustrations, prompting the AI with: \textit{``Can you generate images of a cute rabbit, frog, hedgehog, bear, duck, dinosaur, cat, and puppy.''}, reflecting growing literacy in directing AI tools as design collaborators. T17c used Fotor for AI-based photo generation, whereas T11c turned to Adobe Firefly to create a storyboard thumbnail, though no formal citation was included in this case. Several teams used AI tools explicitly for branding and UI/UX elements. T19c combined ChatGPT and DeepAI to generate logo concepts, \textit{``experimenting with various prompts to achieve the desired visual.''} T5c used Brand Crowd for logo creation and color scheme development, integrating AI into visual identity design. UI-focused tools like Uizard AI were also prevalent. T14a credited Uizard AI directly in a caption: \textit{``Mockup created entirely by the Uizard AI''}.


Our analysis of the documentation phase showed that the students used AI primarily to support writing tasks, including drafting, rewording, summarizing, and organizing content. Our analysis showed that the students often used ChatGPT to help create initial drafts of documentation components. For instance, T26c noted in their reflection that they \textit{``used ChatGPT to help [them] build the structure for [their] documentation,''} highlighting the tool's function in organizing ideas before deeper content development. A dominant theme in the responses was the use of ChatGPT for summarization and improving clarity. T9b described a comprehensive application of AI throughout the documentation phase: \textit{``It helped craft the app's introduction and design rationale and improve sentence phrasing throughout the project documentation,''} emphasizing AI's role as a writing assistant across the entire process. Several teams used AI tools explicitly for rewording and enhancing textual fluency. T13d reported that AI was used \textit{``to help us reword sentences in documentation,''} reflecting the tool's application in stylistic and structural refinement. 



Furthermore, students used AI tools for generating interview protocols, analyzing qualitative data, and video scripting. T12a noted using Microsoft Co-Pilot to \textit{``generate questions for [their] User Testing Protocol,''} while T15b reflected on iteratively improving their work: \textit{``I gave ChatGPT our questions... It gave a few suggestions, and I adjusted the scope from there.''} T31a described using AI \textit{``as a means of generating insights into actionable takeaways,''} showing early use of AI-supported thematic analysis. For scripting, T28 chose to use AI-generated content for their final video, stating they \textit{``decided to go with the ChatGPT version because of its quality,''} pointing to growing trust in AI output for key deliverables.


\subsubsection{Pattern of AI usage and citations}

Using content analysis, 
we categorized AI citation practices into four categories: (1) \textbf{Explicit citation}, where students clearly stated the name of the tool, version or date, and a link to the tool, similar to standard citation guidelines with in-text citation and full reference; (2) \textbf{General Mention}, where students referenced the name of the AI tools without specifying the version or date; (3) \textbf{Indirect mention or lack of explicit citation}, where students did not cite AI use within the text but acknowledged it elsewhere, such as in a separate section discussing AI involvement in their work; and (4) \textbf{No citation mentioned}, where either AI tools were not used, or their use was not mentioned anywhere in the submitted documentation. Among the 35 teams analyzed, five teams fell into the final category, having made no reference to AI use throughout their project. The table~\ref{tab:AI_citation_type} highlights this information excluding the ``no citation'' category.

\textbf{Explicit Citations:} 
These instances of citations included full tool names, version numbers or dates, direct URLs, and formal in-text citation with accompanying bibliographic references. Examples of this include:
\begin{itemize}
    \item Teams T01 and T25a, who cited OpenAI's ChatGPT with versioning and hyperlinks, discussed their use within the literature review and ideation phases.
    \item Teams T24 and T2 offered detailed references, including the exact version and date of the tool used, accompanied by explanations in separate AI usage sections.
    \item For image generation, teams such as T19c and T7c provided explicit citations for tools like DeepAI and Fotor.
\end{itemize}
These students often included prompts, AI outputs, and reflections (as asked) on how the generated content was integrated or modified, indicating a critical and transparent approach to the AI tool usage.

\textbf{General Mention:} Several student submissions referenced AI tools by name (e.g., ChatGPT, Claude 3, Microsoft Co-pilot) without further specification, such as version or access date. This was common in cases like T19b, T5a, and T16c, where students mentioned tools in-text as part of their design rationale or ideation process, but did not provide complete bibliographic citations. In some cases, these general mentions were accompanied by detailed descriptions of how the tools were used in the reflection section (e.g., for generating app names, conducting interviews, or refining documentation), but without formal attribution. This suggests an awareness of the tool's role, but not necessarily of academic citation norms regarding digital tools.

\textbf{Indirect mention or lack of explicit citation:} This category consisted of students who mentioned AI use in reflective sections, appendices, or AI-specific methodology notes, but did not explicitly cite the tools in the main body of their documentation. These students often acknowledged the contributions of tools like ChatGPT, Claude 3, or Adobe Firefly in shaping design outcomes (e.g., generating interview questions, summarizing articles, creating visual assets) but omitted versioning, dates, or formal references.
For example, teams such as T22c, T23b, and T26b discussed the utility of AI tools in shaping their design processes, with references embedded in reflective passages or appendices rather than in-text citations. In image-based work, teams like T11 acknowledged using Adobe Firefly in a ``separate section for AI citation,'' without in-text attribution. Further, a few teams used AI-generated content (e.g., icons, UI mock-ups) and provided only visual or caption-based acknowledgments (e.g., ``icons created using AI''), without naming the specific tools or providing proper references.

\section{Discussion/Provocations}

\subsection{AI Citation as a Pedagogical Opportunity, Not Just a Compliance Task}

Our findings showed that design students are increasingly incorporating generative AI tools across multiple phases of the design process, including brainstorming, secondary research, prototyping, and design documentation. However, the increasing use of AI tools has highlighted significant variability and inconsistency in how students cite the use of AI within their project documentation. We observed that the predominant citation practice adopted by the students was the general mention approach (51.43\%), where most of the students simply mentioned that they used AI without providing in-depth details necessary for examining traces of their design judgment and thinking.  
Sandhaus et al. \cite{sandhaus_student_2024} observed that the absence of AI citation is not always driven by malicious intent but often stems from unclear expectations and a prevailing view of AI as a general utility (such as searching for information on Google Search) rather than a design collaborator. Similarly, He et al. and Wang et al. \cite{he_which_2025, wang_they_2024} highlighted that some students hesitate to disclose AI usage out of concern that doing so may compromise the perceived authenticity or value of their work. 
These challenges, highlighted in prior works and also from the findings of our study, highlight the need to provide sufficient guidance to design students as they incorporate AI into their design process.

From a pedagogical standpoint, we argue that AI citation should be reimagined not as a mere compliance exercise but as a reflective practice that enhances learning. Studio-based design education already centers on process, iteration, and critique, making it a natural environment for encouraging reflective documentation. Within these settings, AI citations can serve as artifacts of design thinking or " design judgment traces" that capture how students utilized AI tools and the reasons behind their use. Encouraging students to reflect on the role that AI played in their workflows allows instructors to better understand the design thinking of students. This reflection will also help students develop a stronger awareness of their own design decisions when using AI in their projects \cite{parsons_separating_2022, parsons_developing_2023}. 

To do this effectively, however, students must engage in metacognitive reflection, documenting not only that they used AI but also how and why. Tankelevitch et al. \cite{tankelevitch_metacognitive_2024} argue that prompting, evaluating, and integrating Generative AI into workflows requires users to actively monitor and control their cognition, adjusting their goals and strategies dynamically. Similarly, Wu et al. \cite{wu_impact_2025} remarked that metacognitive scaffolding significantly enhances the cognitive proficiency of learners in Generative AI-supported environments. This means that when students reflect on why they chose a specific prompt, how they evaluated AI responses, and how they iterated based on those outcomes, they are engaging in a deeper learning process that supports both design development and self-awareness.

These findings motivate us to ask these questions:
\begin{itemize} 
    \item How might we design AI citation practices that support metacognitive development rather than inhibit it through rigid formatting?  
    \item What scaffolds or pedagogical cues can support students in documenting not just tool usage but the thinking behind it? 
\end{itemize}




\subsection{Practical Challenges and Approaches to Citing AI Sources}
Despite its pedagogical potential, citing AI-generated content presents unresolved practical challenges. Traditional citation frameworks like APA and MLA were designed for static, human-authored, and retrievable sources. These frameworks lack the flexibility to accommodate the dynamic, ephemeral, and versioned nature of generative AI outputs \cite{alea_albada_giving_2024, cotton_chatting_2024}. As our study reveals, the citation practices of students varied widely, ranging from detailed metadata to vague mentions or total omission, highlighting a disconnect between existing citation norms and the reality of citing student–AI collaboration in studio-based design education.

This disconnect invites us to rethink what citation could look like in design education. Rather than forcing AI attribution into rigid academic molds, we argue for a dual approach, building from formal citation conventions while incorporating informal, reflective practices that make AI usage visible and pedagogically meaningful.

\textbf{Formal citation} of in-text references with metadata such as tool name, model version, access date, and links mirrors academic conventions where reproducibility is paramount. However, these alone are not enough for students to be transparent about their reasoning for incorporating AI into their design process.

\textbf{Informal citation}, by contrast, is often more reflective of studio-based HCI education. We saw that students who cited AI use through appendices, reflective writing, and visual annotations captured valuable insights—when AI tools were used, how prompts and outputs evolved, and how design decisions were shaped in response. These less standardized disclosures revealed the nuance of student–AI interaction and offered rich pedagogical value.

We propose educators to reframe citation as a form of reflective design documentation—more akin to sketching, critique, or journaling—rather than a compliance mechanism. AI attribution could then be tailored to include when in the design process AI was used (e.g., research, brainstorming, prototyping), what type of content it generated (e.g., text, visuals, interview scripts), how the tool was engaged (e.g., prompt iterations, critiques, refinements), and why the AI-generated content was used, adapted, or discarded.


In this framing, the citation of AI sources becomes a dynamic artifact of student reasoning, especially valuable in iterative, process-driven design work. While formal citation methods are still necessary, a hybrid attribution, one that integrates reflection, process annotations, and contextual documentation, would more accurately capture the evolving role of AI in student–AI collaboration.
These reflections raise the following provocations for HCI educators: \begin{itemize} 
    \item How might we design citation models that reflect both process and outcomes in student–AI collaborations? 
    \item What kinds of informal documentation—such as visual annotations or design journals—can meaningfully supplement formal citations? 
\end{itemize}

\section{Limitations}


This study was conducted within a single undergraduate UX design course, limiting the generalizability of our findings across different institutions, disciplines, or course formats. While the course involved multiple instructors and a sizable sample, students’ prior exposure to AI tools, citation practices, and academic writing varied, which may have shaped their engagement with citation tasks. Additionally, since AI usage was optional and citation guidelines were intentionally minimal, students’ documentation may not fully reflect the scope or depth of AI integration. Our analysis also focused on written artifacts and did not include real-time observations or interviews, which could have offered deeper insight into their citation decision-making processes. We further acknowledge that we did not seek to address the ethical implications of AI usage in student design projects. Future research should consider longitudinal or cross-institutional studies, as well as mixed-method approaches, to build a broader understanding of AI citation in design education. 

\section{Conclusion}
Through this work, we foreground that current AI citation practices are inadequate for capturing the complexities of student–AI collaboration within HCI Design education. We advocate for repositioning AI citation as a pedagogical tool that supports design reflection, judgment, and critical engagement. When AI contribution statements and phase-based documentation models are integrated into design cognition, they can provide visibility into students' design thinking, promote responsible AI use, and uphold the integrity of design assessment. Through this work, we provoke and explore ways design education can adapt to technological shifts while remaining grounded in its core values of process, accountability, and reflective practice. In doing so, we invite educators to consider not only how citation practices should evolve, but also how they can mitigate potential stigma, clarify authorship, and ultimately support student learning. These provocations push us to reimagine AI citation as more than a compliance task—as a meaningful reflection of design process and collaboration.

\bibliographystyle{ACM-Reference-Format}
\bibliography{references}


\begin{thebibliography}{30}


\ifx \showCODEN    \undefined \def \showCODEN     #1{\unskip}     \fi
\ifx \showDOI      \undefined \def \showDOI       #1{#1}\fi
\ifx \showISBNx    \undefined \def \showISBNx     #1{\unskip}     \fi
\ifx \showISBNxiii \undefined \def \showISBNxiii  #1{\unskip}     \fi
\ifx \showISSN     \undefined \def \showISSN      #1{\unskip}     \fi
\ifx \showLCCN     \undefined \def \showLCCN      #1{\unskip}     \fi
\ifx \shownote     \undefined \def \shownote      #1{#1}          \fi
\ifx \showarticletitle \undefined \def \showarticletitle #1{#1}   \fi
\ifx \showURL      \undefined \def \showURL       {\relax}        \fi
\providecommand\bibfield[2]{#2}
\providecommand\bibinfo[2]{#2}
\providecommand\natexlab[1]{#1}
\providecommand\showeprint[2][]{arXiv:#2}

\bibitem[Alea~Albada and Woods(2024)]%
        {alea_albada_giving_2024}
\bibfield{author}{\bibinfo{person}{Nicole Alea~Albada} {and} \bibinfo{person}{Vanessa~E. Woods}.} \bibinfo{year}{2024}\natexlab{}.
\newblock \showarticletitle{Giving {Credit} {Where} {Credit} is {Due}: {An} {Artificial} {Intelligence} {Contribution} {Statement} for {Research} {Methods} {Writing} {Assignments}}.
\newblock \bibinfo{journal}{\emph{Teaching of Psychology}} (\bibinfo{date}{June} \bibinfo{year}{2024}), \bibinfo{pages}{00986283241259750}.
\newblock
\showISSN{0098-6283, 1532-8023}
\urldef\tempurl%
\url{https://doi.org/10.1177/00986283241259750}
\showDOI{\tempurl}


\bibitem[Borg(2000)]%
        {borg_citation_2000}
\bibfield{author}{\bibinfo{person}{Erik Borg}.} \bibinfo{year}{2000}\natexlab{}.
\newblock \showarticletitle{Citation practices in academic writing}.
\newblock \bibinfo{journal}{\emph{Patterns and perspectives: Insights into EAP writing practice}} (\bibinfo{date}{Jan.} \bibinfo{year}{2000}), \bibinfo{pages}{27--45}.
\newblock


\bibitem[Chan and Colloton(2024)]%
        {chan_generative_2024}
\bibfield{author}{\bibinfo{person}{Cecilia Ka~Yuk Chan} {and} \bibinfo{person}{Tom Colloton}.} \bibinfo{year}{2024}\natexlab{}.
\newblock \bibinfo{booktitle}{\emph{Generative {AI} in {Higher} {Education}: {The} {ChatGPT} {Effect}} (\bibinfo{edition}{1} ed.)}.
\newblock \bibinfo{publisher}{Routledge}, \bibinfo{address}{London}.
\newblock
\showISBNx{978-1-00-345902-6}
\urldef\tempurl%
\url{https://doi.org/10.4324/9781003459026}
\showDOI{\tempurl}


\bibitem[Chellappa and Luximon(2024)]%
        {chellappa_understanding_2024}
\bibfield{author}{\bibinfo{person}{Vigneshkumar Chellappa} {and} \bibinfo{person}{Yan Luximon}.} \bibinfo{year}{2024}\natexlab{}.
\newblock \showarticletitle{Understanding the perception of design students towards {ChatGPT}}.
\newblock \bibinfo{journal}{\emph{Computers and Education: Artificial Intelligence}}  \bibinfo{volume}{7} (\bibinfo{date}{Dec.} \bibinfo{year}{2024}), \bibinfo{pages}{100281}.
\newblock
\showISSN{2666-920X}
\urldef\tempurl%
\url{https://doi.org/10.1016/j.caeai.2024.100281}
\showDOI{\tempurl}


\bibitem[Chen et~al\mbox{.}(2021)]%
        {chen_probing_2021}
\bibfield{author}{\bibinfo{person}{Ricky Chen}, \bibinfo{person}{Mychajlo Demko}, \bibinfo{person}{Daragh Byrne}, {and} \bibinfo{person}{Marti Louw}.} \bibinfo{year}{2021}\natexlab{}.
\newblock \showarticletitle{Probing {Documentation} {Practices}: {Reflecting} on {Students}’ {Conceptions}, {Values}, and {Experiences} with {Documentation} in {Creative} {Inquiry}}. In \bibinfo{booktitle}{\emph{Creativity and {Cognition}}}. \bibinfo{publisher}{ACM}, \bibinfo{address}{Virtual Event Italy}, \bibinfo{pages}{1--14}.
\newblock
\showISBNx{978-1-4503-8376-9}
\urldef\tempurl%
\url{https://doi.org/10.1145/3450741.3465391}
\showDOI{\tempurl}


\bibitem[Cotton et~al\mbox{.}(2024)]%
        {cotton_chatting_2024}
\bibfield{author}{\bibinfo{person}{Debby R.~E. Cotton}, \bibinfo{person}{Peter~A. Cotton}, {and} \bibinfo{person}{J.~Reuben Shipway}.} \bibinfo{year}{2024}\natexlab{}.
\newblock \showarticletitle{Chatting and cheating: {Ensuring} academic integrity in the era of {ChatGPT}}.
\newblock \bibinfo{journal}{\emph{Innovations in Education and Teaching International}} \bibinfo{volume}{61}, \bibinfo{number}{2} (\bibinfo{date}{March} \bibinfo{year}{2024}), \bibinfo{pages}{228--239}.
\newblock
\showISSN{1470-3297, 1470-3300}
\urldef\tempurl%
\url{https://doi.org/10.1080/14703297.2023.2190148}
\showDOI{\tempurl}


\bibitem[Dalsgaard et~al\mbox{.}(2013)]%
        {dalsgaard_design_2013}
\bibfield{author}{\bibinfo{person}{Peter Dalsgaard}, \bibinfo{person}{Christian Dindler}, {and} \bibinfo{person}{Jonas Fritsch}.} \bibinfo{year}{2013}\natexlab{}.
\newblock \showarticletitle{Design argumentation in academic design education}.
\newblock \bibinfo{journal}{\emph{Nordes}} \bibinfo{volume}{1}, \bibinfo{number}{5} (\bibinfo{date}{June} \bibinfo{year}{2013}).
\newblock
\showISSN{1604-9705}
\urldef\tempurl%
\url{https://archive.nordes.org/index.php/n13/article/view/331}
\showURL{%
\tempurl}
\newblock
\shownote{Number: 5}.


\bibitem[Dannels(2005)]%
        {dannels_performing_2005}
\bibfield{author}{\bibinfo{person}{Deanna~P. Dannels}.} \bibinfo{year}{2005}\natexlab{}.
\newblock \showarticletitle{Performing {Tribal} {Rituals}: {A} {Genre} {Analysis} of “{Crits}” in {Design} {Studios}}.
\newblock \bibinfo{journal}{\emph{Communication Education}} \bibinfo{volume}{54}, \bibinfo{number}{2} (\bibinfo{date}{April} \bibinfo{year}{2005}), \bibinfo{pages}{136--160}.
\newblock
\showISSN{0363-4523}
\urldef\tempurl%
\url{https://doi.org/10.1080/03634520500213165}
\showDOI{\tempurl}
\newblock
\shownote{Publisher: NCA Website \_eprint: https://doi.org/10.1080/03634520500213165}.


\bibitem[Elo and Kyngäs(2008)]%
        {elo_qualitative_2008}
\bibfield{author}{\bibinfo{person}{Satu Elo} {and} \bibinfo{person}{Helvi Kyngäs}.} \bibinfo{year}{2008}\natexlab{}.
\newblock \showarticletitle{The qualitative content analysis process}.
\newblock \bibinfo{journal}{\emph{Journal of Advanced Nursing}} \bibinfo{volume}{62}, \bibinfo{number}{1} (\bibinfo{year}{2008}), \bibinfo{pages}{107--115}.
\newblock
\showISSN{1365-2648}
\urldef\tempurl%
\url{https://doi.org/10.1111/j.1365-2648.2007.04569.x}
\showDOI{\tempurl}
\newblock
\shownote{\_eprint: https://onlinelibrary.wiley.com/doi/pdf/10.1111/j.1365-2648.2007.04569.x}.


\bibitem[Fazilatfar et~al\mbox{.}(2018)]%
        {fazilatfar_investigation_2018}
\bibfield{author}{\bibinfo{person}{Ali~M. Fazilatfar}, \bibinfo{person}{S.~E. Elhambakhsh}, {and} \bibinfo{person}{Hamid Allami}.} \bibinfo{year}{2018}\natexlab{}.
\newblock \showarticletitle{An {Investigation} of the {Effects} of {Citation} {Instruction} to {Avoid} {Plagiarism} in {EFL} {Academic} {Writing} {Assignments}}.
\newblock \bibinfo{journal}{\emph{SAGE Open}} \bibinfo{volume}{8}, \bibinfo{number}{2} (\bibinfo{date}{April} \bibinfo{year}{2018}), \bibinfo{pages}{2158244018769958}.
\newblock
\showISSN{2158-2440}
\urldef\tempurl%
\url{https://doi.org/10.1177/2158244018769958}
\showDOI{\tempurl}
\newblock
\shownote{Publisher: SAGE Publications}.


\bibitem[Forman and Damschroder(2007)]%
        {forman_qualitative_2007}
\bibfield{author}{\bibinfo{person}{Jane Forman} {and} \bibinfo{person}{Laura Damschroder}.} \bibinfo{year}{2007}\natexlab{}.
\newblock \showarticletitle{Qualitative {Content} {Analysis}}.
\newblock In \bibinfo{booktitle}{\emph{Empirical {Methods} for {Bioethics}: {A} {Primer}}}. Vol.~\bibinfo{volume}{11}. \bibinfo{publisher}{Emerald Group Publishing Limited}, \bibinfo{pages}{39--62}.
\newblock
\showISBNx{978-0-7623-1266-5}
\urldef\tempurl%
\url{https://doi.org/10.1016/S1479-3709(07)11003-7}
\showDOI{\tempurl}
\newblock
\shownote{ISSN: 1479-3709}.


\bibitem[Gray(2018)]%
        {gray_narrative_2018}
\bibfield{author}{\bibinfo{person}{Colin~M. Gray}.} \bibinfo{year}{2018}\natexlab{}.
\newblock \showarticletitle{Narrative {Qualities} of {Design} {Argumentation}}.
\newblock In \bibinfo{booktitle}{\emph{Educational {Technology} and {Narrative}: {Story} and {Instructional} {Design}}}, \bibfield{editor}{\bibinfo{person}{Brad Hokanson}, \bibinfo{person}{Gregory Clinton}, {and} \bibinfo{person}{Karen Kaminski}} (Eds.). \bibinfo{publisher}{Springer International Publishing}, \bibinfo{address}{Cham}, \bibinfo{pages}{51--64}.
\newblock
\showISBNx{978-3-319-69914-1}
\urldef\tempurl%
\url{https://doi.org/10.1007/978-3-319-69914-1_5}
\showDOI{\tempurl}


\bibitem[He et~al\mbox{.}(2025)]%
        {he_which_2025}
\bibfield{author}{\bibinfo{person}{Jessica He}, \bibinfo{person}{Stephanie Houde}, {and} \bibinfo{person}{Justin~D. Weisz}.} \bibinfo{year}{2025}\natexlab{}.
\newblock \bibinfo{title}{Which {Contributions} {Deserve} {Credit}? {Perceptions} of {Attribution} in {Human}-{AI} {Co}-{Creation}}.
\newblock
\newblock
\urldef\tempurl%
\url{https://doi.org/10.48550/arXiv.2502.18357}
\showDOI{\tempurl}
\newblock
\shownote{arXiv:2502.18357 [cs]}.


\bibitem[Hosseini et~al\mbox{.}(2023)]%
        {hosseini_ethics_2023}
\bibfield{author}{\bibinfo{person}{Mohammad Hosseini}, \bibinfo{person}{David~B Resnik}, {and} \bibinfo{person}{Kristi Holmes}.} \bibinfo{year}{2023}\natexlab{}.
\newblock \showarticletitle{The ethics of disclosing the use of artificial intelligence tools in writing scholarly manuscripts}.
\newblock \bibinfo{journal}{\emph{Research Ethics}} \bibinfo{volume}{19}, \bibinfo{number}{4} (\bibinfo{date}{Oct.} \bibinfo{year}{2023}), \bibinfo{pages}{449--465}.
\newblock
\showISSN{1747-0161, 2047-6094}
\urldef\tempurl%
\url{https://doi.org/10.1177/17470161231180449}
\showDOI{\tempurl}


\bibitem[Hsieh and Shannon(2005)]%
        {hsieh_three_2005}
\bibfield{author}{\bibinfo{person}{Hsiu-Fang Hsieh} {and} \bibinfo{person}{Sarah~E. Shannon}.} \bibinfo{year}{2005}\natexlab{}.
\newblock \showarticletitle{Three {Approaches} to {Qualitative} {Content} {Analysis}}.
\newblock \bibinfo{journal}{\emph{Qualitative Health Research}} \bibinfo{volume}{15}, \bibinfo{number}{9} (\bibinfo{date}{Nov.} \bibinfo{year}{2005}), \bibinfo{pages}{1277--1288}.
\newblock
\showISSN{1049-7323}
\urldef\tempurl%
\url{https://doi.org/10.1177/1049732305276687}
\showDOI{\tempurl}
\newblock
\shownote{Publisher: SAGE Publications Inc}.


\bibitem[Istrate(2023)]%
        {istrate_how_2023}
\bibfield{author}{\bibinfo{person}{Olimpius Istrate}.} \bibinfo{year}{2023}\natexlab{}.
\newblock \showarticletitle{How to cite {ChatGPT} and {AI} products}.
\newblock \bibinfo{journal}{\emph{Revista de Pedagogie Digitala}} \bibinfo{volume}{2}, \bibinfo{number}{1} (\bibinfo{year}{2023}), \bibinfo{pages}{3--8}.
\newblock
\showISSN{30082013}
\urldef\tempurl%
\url{https://doi.org/10.61071/RPD.2347}
\showDOI{\tempurl}


\bibitem[Lanning(2016)]%
        {lanning_modern_2016}
\bibfield{author}{\bibinfo{person}{Scott Lanning}.} \bibinfo{year}{2016}\natexlab{}.
\newblock \showarticletitle{A modern, simplified citation style and student response}.
\newblock \bibinfo{journal}{\emph{Reference Services Review}} \bibinfo{volume}{44}, \bibinfo{number}{1} (\bibinfo{date}{Feb.} \bibinfo{year}{2016}), \bibinfo{pages}{21--37}.
\newblock
\showISSN{0090-7324}
\urldef\tempurl%
\url{https://doi.org/10.1108/RSR-10-2015-0045}
\showDOI{\tempurl}


\bibitem[Lee et~al\mbox{.}(2025)]%
        {lee_when_2025}
\bibfield{author}{\bibinfo{person}{Seo-young Lee}, \bibinfo{person}{Matthew Law}, {and} \bibinfo{person}{Guy Hoffman}.} \bibinfo{year}{2025}\natexlab{}.
\newblock \showarticletitle{When and {How} to {Use} {AI} in the {Design} {Process}? {Implications} for {Human}-{AI} {Design} {Collaboration}}.
\newblock \bibinfo{journal}{\emph{International Journal of Human–Computer Interaction}} (\bibinfo{date}{Jan.} \bibinfo{year}{2025}).
\newblock
\showISSN{1044-7318}
\urldef\tempurl%
\url{https://www.tandfonline.com/doi/full/10.1080/10447318.2024.2353451}
\showURL{%
\tempurl}
\newblock
\shownote{Publisher: Taylor \& Francis}.


\bibitem[Lively et~al\mbox{.}(2023)]%
        {lively_integrating_2023}
\bibfield{author}{\bibinfo{person}{Jason Lively}, \bibinfo{person}{James Hutson}, {and} \bibinfo{person}{Elizabeth Melick}.} \bibinfo{year}{2023}\natexlab{}.
\newblock \showarticletitle{Integrating {AI}-{Generative} {Tools} in {Web} {Design} {Education}: {Enhancing} {Student} {Aesthetic} and {Creative} {Copy} {Capabilities} {Using} {Image} and {Text}-{Based} {AI} {Generators}}.
\newblock \bibinfo{journal}{\emph{DS Journal of Artificial Intelligence and Robotics}} \bibinfo{volume}{1}, \bibinfo{number}{1} (\bibinfo{date}{Sept.} \bibinfo{year}{2023}), \bibinfo{pages}{23--36}.
\newblock
\showISSN{25839926}
\urldef\tempurl%
\url{https://doi.org/10.59232/AIR-V1I1P103}
\showDOI{\tempurl}


\bibitem[McAdoo(2023)]%
        {mcadoo_how_2023}
\bibfield{author}{\bibinfo{person}{Timothy McAdoo}.} \bibinfo{year}{2023}\natexlab{}.
\newblock \bibinfo{title}{How to cite {ChatGPT}}.
\newblock
\newblock
\urldef\tempurl%
\url{https://apastyle.apa.org/blog/how-to-cite-chatgpt}
\showURL{%
\tempurl}


\bibitem[Papachristos et~al\mbox{.}(2024)]%
        {papachristos_integrating_2024}
\bibfield{author}{\bibinfo{person}{Eleftherios Papachristos}, \bibinfo{person}{Yavuz Inal}, \bibinfo{person}{Carlos~Vicient Monllaó}, \bibinfo{person}{Eivind~Arnstein Johansen}, {and} \bibinfo{person}{Mari Hermansen}.} \bibinfo{year}{2024}\natexlab{}.
\newblock \bibinfo{title}{Integrating {AI} into {Design} {Ideation}: {Assessing} {ChatGPT}’s {Role} in {Human}-{Centered} {Design} {Education}}.
\newblock
\newblock
\urldef\tempurl%
\url{https://doi.org/10.36227/techrxiv.171656320.09963657/v1}
\showDOI{\tempurl}


\bibitem[Parsons and Gray(2022)]%
        {parsons_separating_2022}
\bibfield{author}{\bibinfo{person}{Paul~C Parsons} {and} \bibinfo{person}{Colin~M Gray}.} \bibinfo{year}{2022}\natexlab{}.
\newblock \showarticletitle{Separating {Grading} and {Feedback} in {UX} {Design} {Studios}}. In \bibinfo{booktitle}{\emph{{EduCHI}'22: 4th {Annual} {Symposium} on {HCI} {Education}}}. \bibinfo{address}{New Orleans, LA}, \bibinfo{pages}{9}.
\newblock


\bibitem[Parsons et~al\mbox{.}(2023)]%
        {parsons_developing_2023}
\bibfield{author}{\bibinfo{person}{Paul~C Parsons}, \bibinfo{person}{Prakash~Chandra Shukla}, \bibinfo{person}{Ali Baigelenov}, {and} \bibinfo{person}{Colin~M Gray}.} \bibinfo{year}{2023}\natexlab{}.
\newblock \showarticletitle{Developing {Framing} {Judgment} {Ability}: {Student} {Perceptions} from a {Graduate} {UX} {Design} {Program}}. In \bibinfo{booktitle}{\emph{Proceedings of the 5th {Annual} {Symposium} on {HCI} {Education}}}. \bibinfo{publisher}{ACM}, \bibinfo{address}{Hamburg Germany}, \bibinfo{pages}{23--32}.
\newblock
\showISBNx{9798400707377}
\urldef\tempurl%
\url{https://doi.org/10.1145/3587399.3587401}
\showDOI{\tempurl}


\bibitem[Sandhaus et~al\mbox{.}(2024)]%
        {sandhaus_student_2024}
\bibfield{author}{\bibinfo{person}{Hauke Sandhaus}, \bibinfo{person}{Quiquan Gu}, \bibinfo{person}{Maria~Teresa Parreira}, {and} \bibinfo{person}{Wendy Ju}.} \bibinfo{year}{2024}\natexlab{}.
\newblock \bibinfo{title}{Student {Reflections} on {Self}-{Initiated} {GenAI} {Use} in {HCI} {Education}}.
\newblock
\newblock
\urldef\tempurl%
\url{http://arxiv.org/abs/2410.14048}
\showURL{%
\tempurl}
\newblock
\shownote{arXiv:2410.14048 [cs]}.


\bibitem[Shi et~al\mbox{.}(2023)]%
        {shi_understanding_2023}
\bibfield{author}{\bibinfo{person}{Yang Shi}, \bibinfo{person}{Tian Gao}, \bibinfo{person}{Xiaohan Jiao}, {and} \bibinfo{person}{Nan Cao}.} \bibinfo{year}{2023}\natexlab{}.
\newblock \showarticletitle{Understanding {Design} {Collaboration} {Between} {Designers} and {Artificial} {Intelligence}: {A} {Systematic} {Literature} {Review}}.
\newblock \bibinfo{journal}{\emph{Proceedings of the ACM on Human-Computer Interaction}} \bibinfo{volume}{7}, \bibinfo{number}{CSCW2} (\bibinfo{date}{Sept.} \bibinfo{year}{2023}), \bibinfo{pages}{1--35}.
\newblock
\showISSN{2573-0142}
\urldef\tempurl%
\url{https://doi.org/10.1145/3610217}
\showDOI{\tempurl}


\bibitem[Shukla et~al\mbox{.}(2025)]%
        {shukla_-skilling_2025}
\bibfield{author}{\bibinfo{person}{Prakash Shukla}, \bibinfo{person}{Phuong Bui}, \bibinfo{person}{Sean~S Levy}, \bibinfo{person}{Max Kowalski}, \bibinfo{person}{Ali Baigelenov}, {and} \bibinfo{person}{Paul Parsons}.} \bibinfo{year}{2025}\natexlab{}.
\newblock \showarticletitle{De-skilling, {Cognitive} {Offloading}, and {Misplaced} {Responsibilities}: {Potential} {Ironies} of {AI}-{Assisted} {Design}}. In \bibinfo{booktitle}{\emph{Proceedings of the {Extended} {Abstracts} of the {CHI} {Conference} on {Human} {Factors} in {Computing} {Systems}}} \emph{(\bibinfo{series}{{CHI} {EA} '25})}. \bibinfo{publisher}{Association for Computing Machinery}, \bibinfo{address}{New York, NY, USA}, \bibinfo{pages}{1--7}.
\newblock
\showISBNx{9798400713958}
\urldef\tempurl%
\url{https://doi.org/10.1145/3706599.3719931}
\showDOI{\tempurl}


\bibitem[Shukla et~al\mbox{.}(2024)]%
        {shukla_communication_2024}
\bibfield{author}{\bibinfo{person}{Prakash Shukla}, \bibinfo{person}{Suchismita Naik}, \bibinfo{person}{Ike Obi}, \bibinfo{person}{Phuong Bui}, {and} \bibinfo{person}{Paul Parsons}.} \bibinfo{year}{2024}\natexlab{}.
\newblock \showarticletitle{Communication {Challenges} {Reported} by {UX} {Designers} on {Social} {Media}: {An} {Analysis} of {Subreddit} {Discussions}}. In \bibinfo{booktitle}{\emph{Extended {Abstracts} of the 2024 {CHI} {Conference} on {Human} {Factors} in {Computing} {Systems}}} \emph{(\bibinfo{series}{{CHI} {EA} '24})}. \bibinfo{publisher}{Association for Computing Machinery}, \bibinfo{address}{New York, NY, USA}, \bibinfo{pages}{1--6}.
\newblock
\showISBNx{9798400703317}
\urldef\tempurl%
\url{https://doi.org/10.1145/3613905.3650881}
\showDOI{\tempurl}


\bibitem[Tankelevitch et~al\mbox{.}(2024)]%
        {tankelevitch_metacognitive_2024}
\bibfield{author}{\bibinfo{person}{Lev Tankelevitch}, \bibinfo{person}{Viktor Kewenig}, \bibinfo{person}{Auste Simkute}, \bibinfo{person}{Ava~Elizabeth Scott}, \bibinfo{person}{Advait Sarkar}, \bibinfo{person}{Abigail Sellen}, {and} \bibinfo{person}{Sean Rintel}.} \bibinfo{year}{2024}\natexlab{}.
\newblock \showarticletitle{The {Metacognitive} {Demands} and {Opportunities} of {Generative} {AI}}. In \bibinfo{booktitle}{\emph{Proceedings of the {CHI} {Conference} on {Human} {Factors} in {Computing} {Systems}}}. \bibinfo{publisher}{ACM}, \bibinfo{address}{Honolulu HI USA}, \bibinfo{pages}{1--24}.
\newblock
\showISBNx{9798400703300}
\urldef\tempurl%
\url{https://doi.org/10.1145/3613904.3642902}
\showDOI{\tempurl}


\bibitem[Wang and Fussell(2024)]%
        {wang_they_2024}
\bibfield{author}{\bibinfo{person}{Yadi Wang} {and} \bibinfo{person}{Susan~R. Fussell}.} \bibinfo{year}{2024}\natexlab{}.
\newblock \showarticletitle{They {May} {Have} {Seen} {My} {ChatGPT} {Tab}: {Exploring} {Social} {Perceptions} of {AI}-{Assisted} {Writing} for {ESL} {Students}}. In \bibinfo{booktitle}{\emph{Proceedings of the {Third} {Workshop} on {Intelligent} and {Interactive} {Writing} {Assistants}}}. \bibinfo{publisher}{ACM}, \bibinfo{address}{Honolulu HI USA}, \bibinfo{pages}{13--15}.
\newblock
\showISBNx{9798400710315}
\urldef\tempurl%
\url{https://doi.org/10.1145/3690712.3690717}
\showDOI{\tempurl}


\bibitem[Wu et~al\mbox{.}(2025)]%
        {wu_impact_2025}
\bibfield{author}{\bibinfo{person}{Junqi Wu}, \bibinfo{person}{Wang  , Jiatong}, \bibinfo{person}{Lei  , Shuang}, \bibinfo{person}{Wu  , Feiyan}, \bibinfo{person}{}, {and} \bibinfo{person}{Xingyu Gao}.} \bibinfo{year}{2025}\natexlab{}.
\newblock \showarticletitle{The impact of metacognitive scaffolding on deep learning in a {GenAI}-supported learning environment}.
\newblock \bibinfo{journal}{\emph{Interactive Learning Environments}} \bibinfo{volume}{0}, \bibinfo{number}{0} (\bibinfo{year}{2025}), \bibinfo{pages}{1--18}.
\newblock
\showISSN{1049-4820}
\urldef\tempurl%
\url{https://doi.org/10.1080/10494820.2025.2479162}
\showDOI{\tempurl}


\end{thebibliography}










\end{document}